\documentclass{iopart}
\usepackage[dvips]{graphicx}
\usepackage{wrapfig,epsfig}
\usepackage{epsf}
\usepackage{iopams}
\begin{document}
\baselineskip 14pt
\noindent

\begin{tabbing}
\hspace{11cm}\=\kill
 \> \today.
\end{tabbing}

\vspace{1cm}
\noindent
{\LARGE  The structure of fluid trifluoromethane and methylfluoride}
\vspace{1cm}

\noindent
{\large J.~Neuefeind $^{\mathrm 1}$, H.E.~Fischer $^{\mathrm 2}$, 
W.~Schr\"oer $^{\mathrm 3}$} \\[.5cm]
 
\begin{center} \begin{minipage}[c]{10cm}
$^{\mathrm 1}$ Hamburger Synchrotronstrahlungslabor HASYLAB at DESY, 
Notkestr 85, D-22603 Hamburg Germany\\
$^{\mathrm 2}$ Institut Laue-Langevin, 6 rue Jules Horowitz, BP~156, 
38042 Grenoble Cedex 9,
France \footnote{present address: Laboratoire pour l'Utilisation du 
Rayonnement Electromagn\'etique (LURE), Centre Universitaire Paris-Sud, 
BP~34, 91898 ORSAY cedex, France}\\
$^{\mathrm 3}$ Inst. f. Anorg. und Phys. Chem, Univ. Bremen,  
Leobener Str. NW II, D-28359 Bremen, Germany
\end{minipage}\\[.5cm]

\vspace{1cm}
\begin{abstract}
We present hard X-ray and neutron diffraction measurements on the polar 
fluorocarbons HCF$_3$ and H$_3$CF under supercritical conditions and 
for a range of molecular densities spanning about a factor of ten.
The Levesque-Weiss-Reatto inversion scheme has been used to deduce 
the site-site potentials underlying the measured partial pair 
distribution functions. The orientational correlations between
adjacent fluorocarbon molecules -- which are characterized by  
quite large dipole moments but no tendency to form hydrogen bonds --
are small compared to a highly polar system like fluid hydrogen
chloride.  In fact, the orientational correlations in HCF$_3$ and 
H$_3$CF are found to be nearly as small as those of fluid CF$_4$, 
a fluorocarbon with no dipole moment.
\end{abstract}
\vspace{1cm}
\noindent\par
\end{center}

\section{Introduction}
The understanding of dielectric properties resulting from orientational 
correlations\cite{Buc67,Sch85}, as well as the determination of 
orientational 
correlations from diffraction experiments\cite{Zei82,Sop91}, are long standing 
problems in the physics of molecular fluids.  The simple
fluorocarbons HCF$_3$ and H$_3$CF are very interesting model  
substances in this context, as they posses rather large dipole
 moments ($1.65\cdot 3.336\cdot10^{-30}$Cm), in the 
case of trifluoromethane -- the same as that of the water molecule)
but no tendency to form hydrogen bonds \cite{Franck}. 
The investigation of the structure of the simple fluorocarbons 
H$_3$CF and HCF$_3$ thus enables the study of the structural effect of
the molecular dipole alone.

Although the properties of these fluorocarbons are interesting, 
only very limited structural information is available so far. HCF$_3$ is 
discussed as replacement for chlorinated hydrocarbons as refrigerant since 
it it has no ozone damaging effect, it has a shorter atmospheric lifetime and,
 hence, a lower global warming potential and it presents no toxological risk 
\cite{pires,fermeglia}. HCF$_3$ is discussed for extraction applications 
\cite{zhao} and it has been shown, that the enatioselectivity
of asymmetric catalysis can be controlled by the density of the fluoroform
solvent \cite{wynne}.
The crystal structure of HCF$_3$ has been determined 
by a neutron powder diffraction experiment \cite{Tor96} and the molecular 
geometry by a gas phase electron diffraction study \cite{Kaw78}. 
The only fluorocarbon investigated in the supercritical regime is 
tetrafluoromethane\cite{Wal}, whereas deuterated trifluoromethane has been 
investigated in the liquid regime \cite{MP1,MP2}.  
In both cases the total neutron structure factor has been 
determined. References to the simulation studies performed for 
HCF$_3$ can be found in the recent work by Hloucha {\it et al} \cite{Dei}
beginning with the early work of B\"ohm {\it et al} \cite{boehm}.

The method used here to deduce molecular orientations is based on the 
potential inversion scheme of Levesque, Weis and Reatto \cite{LWR}. From
the result of hard X-ray and neutron diffraction experiments with
isotopic substitution (NDIS)  a site-site 
potential is deduced, which in turn can be used in a NVT-Monte-Carlo 
simulation to obtain the orientational correlations. The determination of the 
potential is facilitated by the ease with which the density of these systems
can be varied, both having a critical point at about room temperature.

\section{Experimental}
We have investigated the structure of the fluid fluorocarbons in a range 
of pressures (28-100\,bar) and temperatures (298-333\,K) around the 
critical points of 58\,bar, 317.8\,K for H$_3$CF and 48.3\,bar, 299.1\,K 
for HCF$_3$ \cite{CRC}. It was aimed to combine the information
of a NDIS experiment \cite{ENE,PSF} and a 
hard X-ray diffraction experiment \cite{Pou}. The neutron and hard X-ray 
experiments were
both carried out using the same sample environment 
(Fig.~\ref{abb-hd-hcf3}) built especially for this experiment.
The mechanical requirements of the pressure cell are moderate, and 
aluminum is a very suitable material for the sample container. Aluminum has a 
quite low scattering power for both neutrons and X-rays, only few powder 
lines due to its cubic structure and shows only little activation in a 
neutron beam.  The samples, DCF$_3$ (98\% D, Cambridge isotopes),
HCF$_3$, a 1:1 mixture HCF$_3$/DCF$_3$ and H$_3$CF (all Linde technical 
gases) can be condensed into the sample container through immersion in 
liquid nitrogen. The cell is then mounted inside a vacuum tank
of the neutron or hard X-ray diffractometer. The sample pressure
inside the mounted cell can be varied via an inert gas line, separated by a 
steel bellow from the sample. The temperature can be varied with a 
small heater at the bottom of the cell. The price of the deuterated
gases requires the reduction of dead volumes inside the cell:
The cell can be filled with $\sim4\,$g sample.
\begin{figure}[htb]
\caption{\label{abb-hd-hcf3}Schematic drawing and picture of the 
pressure set-up\\A: sample container, B: steel bellow, C: pressure sensor, D: 
heater,
E: temperature sensor, F: vent for the pressurizing medium, 
G: sample vent.}
{\small
}

\end{figure}

\begin{figure}[htbp]
\includegraphics[scale=.7]{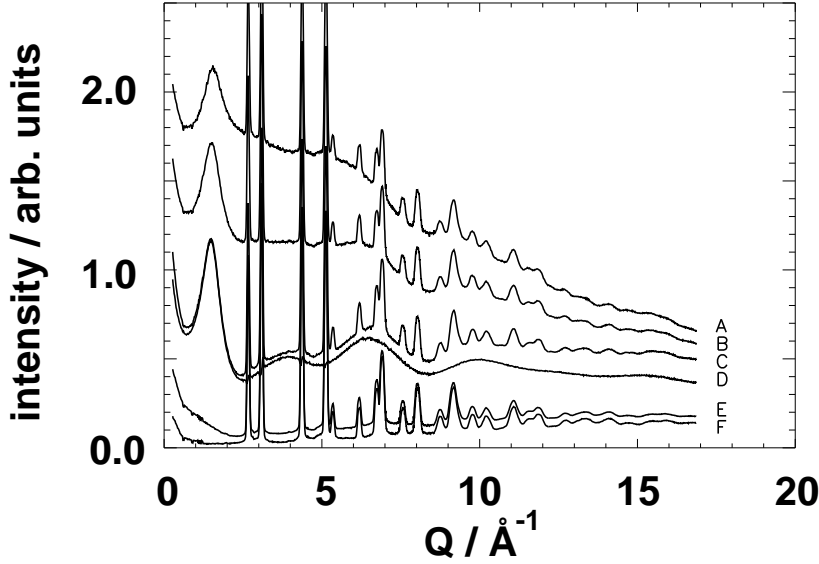}

\caption{Raw data of the neutron diffraction experiment on trifluoromethane
at D4b.
\label{abb-roh-hcf3}.\\
A-C: Sample + cell scattering intensity of HCF$_3$, 
MCF$_3$ (the H/D mixture)
and DCF$_3$ respectively at liquid-like densities (30 $^{\rm o}$C, 100 bar),
D: scattering intensity of C after subtraction of the cell scattering,
E: Sample + cell scattering intensity of DCF$_3$
at gas-like densities (60 $^{\rm o}$C, 32 bar), F: empty aluminum container. 
}
\end{figure}

The neutron diffraction measurements were performed at the 
diffractometer D4b\cite{YB} at the ILL reactor source in Grenoble, 
using a wavelength of 0.7501\,\AA. Representative examples of the raw data 
are shown in
Fig.~\ref{abb-roh-hcf3}, showing that the aluminum cell is clearly a viable
alternative to the more usual vanadium and titanium-zirconium cells. 
The hard X-ray diffraction was performed at the high energy beamline BW5
at the DORIS storage ring at HASYLAB, Hamburg in its set-up for liquid and
amorphous substances \cite{Bou98b}, using a wavelength of 0.1282\,\AA

\section{Data analysis}
\begin{figure}[htbp]
\begin{center}
\includegraphics[scale=.70]{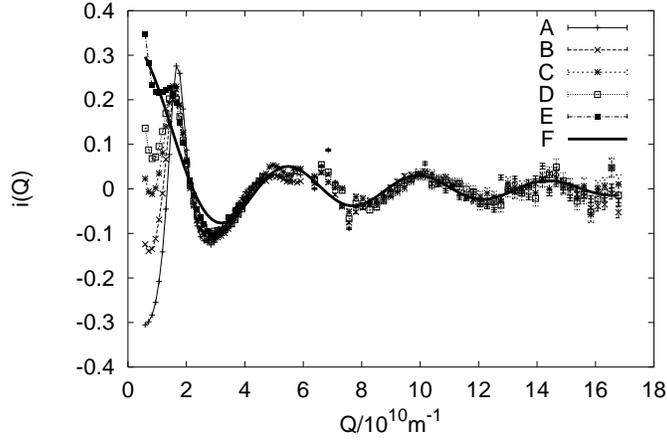}

\rule{0pt}{0pt}
\end{center}
\caption{\label{abb-h3cf-qiq} X-ray structure 
function $i(Q)$
of H$_3$CF at various densities.\\
A: $\rho_{uc}=17.6\cdot 10^{-3}$\,\AA$^{-3}$,
B: $\rho_{uc}=13.0\cdot 10^{-3}$\,\AA$^{-3}$,
C: $\rho_{uc}=10.4\cdot 10^{-3}$\,\AA$^{-3}$,
D: $\rho_{uc}= 7.3\cdot 10^{-3}$\,\AA$^{-3}$,
E: $\rho_{uc}= 2.6\cdot 10^{-3}$\,\AA$^{-3}$,
F: Fit of equation~\ref{eq-debye} to the data with 
$r_{\rm CF}= 1.416(8)$\,\AA\ and  $\gamma_{\rm CF}= 0.050(18)$\,\AA.}
\end{figure}

The data were corrected for systematic effects like detector dead time,
absorption, container scattering, multiple and incoherent scattering, 
using the procedure
described in some detail in \cite{allohol}, and then normalized.
The differential cross sections are 
expressed in terms of the scattering functions $S^{(x)}(Q)$ and $S^{(n)}(Q)$
for the hard X-ray and the neutron cases
\begin{eqnarray}
S^{(n)}(Q)&=&\frac{\left(\frac{d\sigma}{d\Omega}\right)^{(n)}-
\sum_{i}^{N_{uc}} \nu_i b_i^2}{(\sum_{i}^{N_{uc}} \nu_i b_i)^2}+1 
\label{eq-sq-n}\\
S^{(x)}(Q)=i(Q)+1&=&\frac{\left(\frac{d\sigma}{d\Omega}\right)^{(x)}/
\sigma_{el}-\sum_{i}^{N_{uc}} \nu_i f_i^2}{(\sum_{i}^{N_{uc}} \nu_i f_i)^2}
+1\label{eq-sq-x}
\end{eqnarray}
where $\left(\frac{d\sigma}{d\Omega}\right)$ is the coherent 
differential cross section, $b_i$ the coherent scattering lengths \cite{Koe},
$f_i$ the X-ray form factors in the independent atom 
approximation \cite{Hub75}, 
$\sigma_{el}$ the scattering cross section of the free electron, $\nu_i$ the 
stoichiometric coefficient of the atom $i$, and where the sums are extending 
over the number of distinct atoms in the molecule $N_{uc}$, the subscript $uc$
referring to the unit of composition, the molecule.
In Fig.~\ref{abb-h3cf-qiq} the density dependence of the X-ray structure 
function of H$_3$CF is shown. Beyond $Q\sim 2.5$\,\AA$^{-1}$ the interference 
scattering intensity is dominated by the intramolecular contributions.
Fitting of the Debye equation in the range $[4<Q<Q_{max}]$:
\begin{equation}
\label{eq-debye}
i(Q)_{\rm intra} = \sum_{i \neq j} 2 \frac{f_if_j}{(\sum_{i}^{N_{uc}} 
\nu_i f_i)^2} \frac{\sin(Qr_{ij,eq})}{Qr_{ij,eq}}
\exp(-Q^2\gamma_{ij}^2/2.)\, ,
\end{equation}
with $r_{ij,eq}$ the equilibrium distance of the atoms $i$ and $j$ 
within the molecule and $\gamma_{ij}$ the displacement parameter,
leads to $r_{\rm CF}= 1.416(8)$\,\AA\ and  $\gamma_{\rm CF}= 0.050(18)$\,\AA\
independent of density. Likewise, the molecular parameters of trifluoromethane 
were determined and are shown in Table~\ref{tab-molpar-hcf3}.
For trifluoromethane the molecular parameters are 
also independent of the density and in excellent agreement with the gas 
phase values from ref.~\cite{Kaw78}. For the remainder of the article only 
the intermolecular contributions to the structure are considered.
 \begin{table}[b]
\begin{center}
    \begin{tabular}{|@{}l|l@{}l@{}l@{}|}
    \hline
    &  This work, fluid \hspace{.3cm}\,& crystal \cite{Tor96}\hspace{.3cm}\,& 
gas phase 
\cite{Kaw78}\hspace{.3cm}\,\\ \hline
$r_{CF}$&1.327&1.315(4)$^2$&1.3284(31)\\ 
$r_{FF}$&2.153&&\\
$r_{HC}$&1.088&1.111(7)&1.091(14)\\
$r_{HF}$&1.995&&\\
$\gamma_{CF}$&.092&&\\
$\gamma_{FF}$&0.104&&\\
$\gamma_{HC}$&0.112&&\\
$\gamma_{HF}$&0.145&&\\
$\angle HCF$&111.0$^1$&109.77(32)$^2$&110.35\\
$\angle FCF$&108.4$^1$&109.14(43)$^2$&108.58(44)\\ \hline
  \end{tabular}

\rule{0cm}{0cm}\end{center}
\caption{Molecular structure of HCF$_3$.\label{tab-molpar-hcf3}
\\ All distances and displacement parameter in \AA , all angles in 
degree.\\
$^1$ The angles are not refined directly, but determined from the 
maxima of the distance distributions.\\
$^2$ Mean value of intramolecular distances and bond angles non equivalent 
in the crystal.}
\end{table}

The intermolecular scattering contribution is related to the weighted
intermolecular pair distribution functions by a Fourier-sine transformation:
\begin{eqnarray}
r \cdot (g^{(n)}-1) &=& \frac{1}{2\pi^2\rho_{uc}} 
\int Q \cdot (S^{(n)}-1)  
\sin(Qr) dQ \label{eq-ft-n}
\\ 
r \cdot (g^{(x)}-1) &=& \frac{1}{2\pi^2\rho_{uc}} 
\int Q \cdot i(Q) \sin(Qr) dQ \label{eq-ft-x}\:.
\end{eqnarray}
where $\rho_{uc}$ is the density per unit of composition (molecule). 
$g^{(x)}$ and $g^{(n)}$ and are 
weighted sums of the partial (site-site) pair distribution functions (PPDF):

\begin{equation}
g^{(n)}= \sum_{ij} w_{ij} g_{ij}\, \, \, \, \mbox{with}\,\,\,\,\, 
w_{ij}=\frac{\nu_i\nu_j b_i b_j}{(\sum_i \nu_i b_i)^2} 
\end{equation}
and 
\begin{equation}
g^{(x)} = \sum_{ij} \mbox{FT}[w_{ij}(Q)] \otimes g_{ij} \,\, \, \, 
\mbox{ with}\,\,\,\,\,
w_{ij}(Q)=\frac{\nu_i \nu_j f_i(Q) f_j(Q)}{(\sum \nu_i f_i)^2} \: ,   
  \label{eq-ftx}
\end{equation}
where $\mbox{FT}()$ is the Fourier sine transformation and $\otimes$ the 
convolution operation.

Trifluoromethane has six PPDF, and four independent diffraction 
experiments were carried out. Consequently, assuming the 
independence of the structure from the isotopic composition, a set 
of one PPDF (HH) and 
three independent composite partial pair distribution functions (CPPDF) 
can be isolated, each of the CPPDF is the weighted sum of two PPDF. 
The three CPPDF are dominated by the fluorine PPDF, FF, FC and FH, 
while the carbon PPDF, CC and HC do contribute only very little. 
Alternatively the total 
pair distribution function can be split into HH, HX and XX CPPDF, with X 
either C or F. This is the same separation as that used in the case of the 
hydrogenhalides, with X=Cl in that case, thus enabling a proper comparison of 
our results
with measurements of the structure of fluid HCl.
All pair distribution functions are defined such
that $\lim_{r\rightarrow \infty} g(r)=1.$

\section{Potential inversion}
In order to generate a three dimensional picture of the structure from the
pair distribution functions, the potential inversion scheme of Levesque, 
Weiss and Reatto \cite{LWR} (LWR-scheme) was applied. The idea of 
this method is based on the equation:
\begin{equation}
g(r)= \exp \left[ \frac{-v(r)}{kT}  + g(r) - 1 - c(r) + B(r,v)\right]
\label{eq-hd}
\end{equation}
relating the pair distribution function and the pair potential, where
$v(r)$ is the pair potential, $c(r)$ the direct correlation function and 
B(r,v) the bridge function. Starting with a first guess of the potential
$v^{(1)}$, e.g. by neglecting the bridge function, 
a Monte Carlo simulation gives $g(r)^{(1)}$ and $c(r)^{(1)}$ 
belonging to $v^{(1)}$ and 
thus $B(r,v^{(1)})$.
Substituting $B(r,v^{(n-1)})$ for $B(r,v)$ in equation~\ref{eq-hd} gives 
the LWR iteration formula
\begin{eqnarray}
\nonumber &&v^{(n)}/kT=v^{(n-1)}/kT + \mathrm{ln} (g^{(n-1)}/g^{(exp)}) 
\\&&+ c^{(n-1)} - c^{(exp
)}
- g^{(n-1)} + g^{(exp)}\label{eq-iter}
\end{eqnarray}
Schommers \cite{Sch83} proposed a similar iteration scheme where only 
the logarithmic term of Eq.~\ref{eq-iter} is considered. Reatto {\it et al} 
have shown \cite{RLW}, that their algorithm converges 
- under certain conditions much - faster compared to the Schommers scheme.
 Soper's EPMC algorithm \cite{Sop96} uses the same iteration scheme 
as Schommers, 
but contrary to it is formulated for multi-element fluids. 
Likewise, it has been shown by Kahl and Kristufek \cite{Kah} 
that the LWR scheme is applicable to polyatomic systems. The systems 
investigated here are even a step more complicated than the systems 
Kahl and Kristufek used, as the sites are connected by covalent bonds and the 
PPDF are not complete. Thus the HC and the CC site-site potentials were 
kept constant as hard sphere potentials. The method has already been tested 
under these conditions and compared to the results of Reverse Monte-Carlo 
(RMC) \cite{RMC} simulations.
A short account of the comparison between the potential inversion and the 
RMC method has been given elsewhere \cite{phy_b}. 
In the following only the results of the potential 
inversion method are given.

\begin{figure}[htbp]
\begin{center}
\includegraphics[scale=.7]{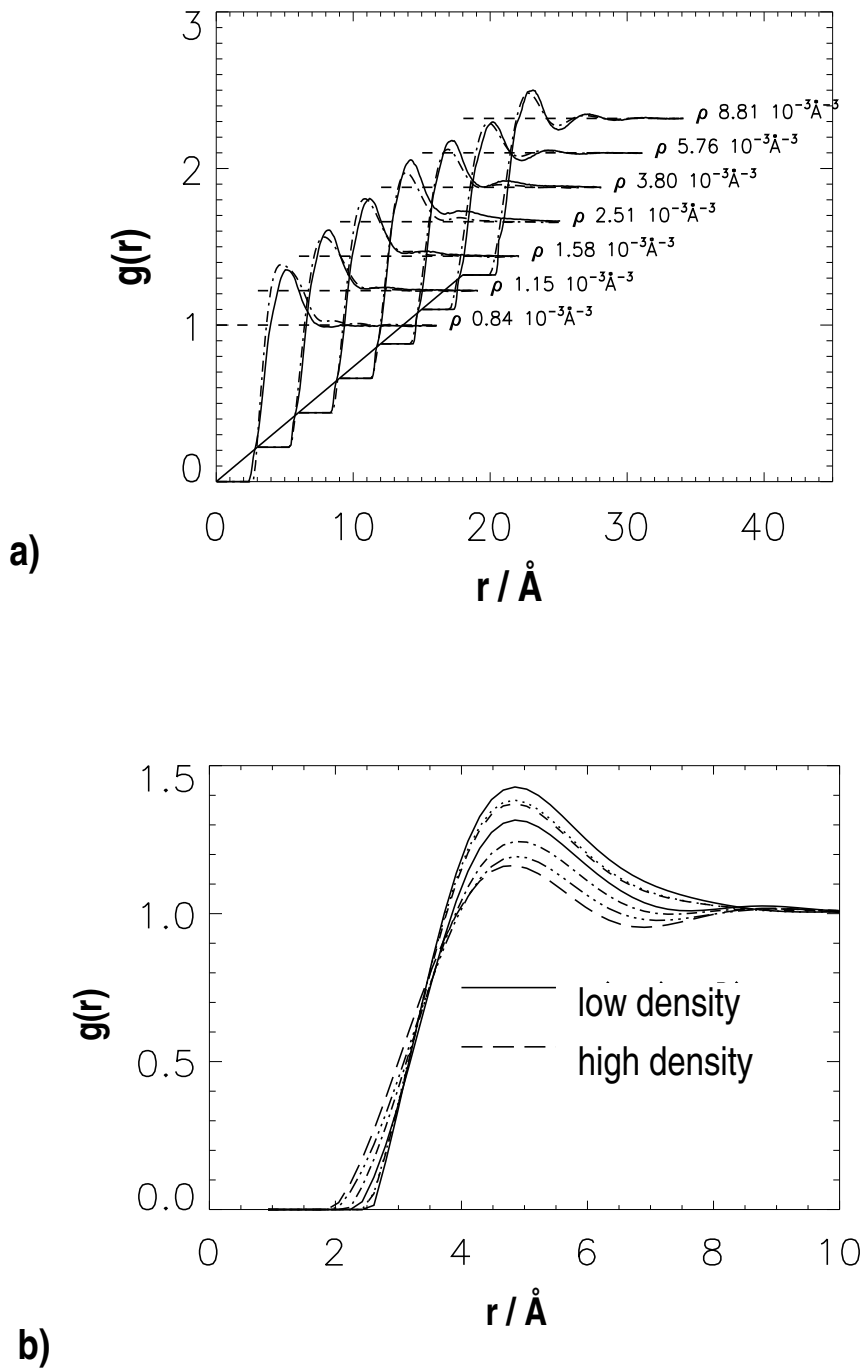}

\rule{0pt}{0pt}\end{center}
\caption{a)Example of the density dependence of the intermolecular part  of
 the $g^{(n)}_{XX)}$ CPPDF of HCF$_3$. The experimental 
$g_{XX}^{(n)}$ is compared with the prediction of
a potential model derived at some specific state point 
(f:$\rho_{uc}$=1.15 10$^{-3}$ \AA$^{-3}$).\\
b) Direct comparison of the experimental $g_{XX}^{(n)}$ at decreasing 
density. 
(same state points as in a)
\label{abb-ddstruck-hcf3}\\
The densities in the figure correspond to the following experimental
 conditions (from high to low density) : a: p=100 bar, T=298K, 
$\rho_{uc}$= 8.8 10$^{-3}$ \AA$^{-3}$,
b: p=100 bar, T=333K, $\rho_{uc}$= 5.8 10$^{-3}$ \AA$^{-3}$, 
c: p=80 bar, T=333K, $\rho_{uc}$= 3.8 10$^{-3}$ \AA$^{-3}$, 
d: p=60 bar, T=333K, $\rho_{uc}$= 2.5 10$^{-3}$ \AA$^{-3}$, 
e: p=50 bar, T=333K, $\rho_{uc}$=1.6 10$^{-3}$ \AA$^{-3}$, 
f: p=40 bar, T=333K, $\rho_{uc}$=1.15 10$^{-3}$ \AA$^{-3}$,
g: p=32 bar, T=333K, $\rho_{uc}$=0.84 10$^{-3}$ \AA$^{-3}$.
}
\end{figure}

One test of the potential inversion scheme is the correct or incorrect 
prediction of the pair distribution function at several different state points 
from a pair potential determined at a specific state point. This is 
demonstrated in Fig.~\ref{abb-ddstruck-hcf3}. This procedure  
also tests whether 
the intermolecular potential can be described as an effective two-body
potential.

\section{Results and discussion}
In Fig.~\ref{abb-ddstruck-hcf3}a the density dependence of $g^{(n)}_{XX}$ of 
trifluoromethane is shown. The density dependence of the other CPPDF is 
similar. At the higher densities $g^{(n)}_{XX}$ shows a typical liquid-like 
behavior and several maxima and minima. The maxima at larger distances
die out when lowering the density, while the height of the first maximum
increases. This is opposite to the behavior of fluid hydrogen chloride 
\cite{And97} or water \cite{Sop97} where the height of the main 
maximum decreases with density.
 The position of the main maximum remains almost unchanged. 

\begin{figure}[htbp]
\begin{center}
\includegraphics[scale=.7]{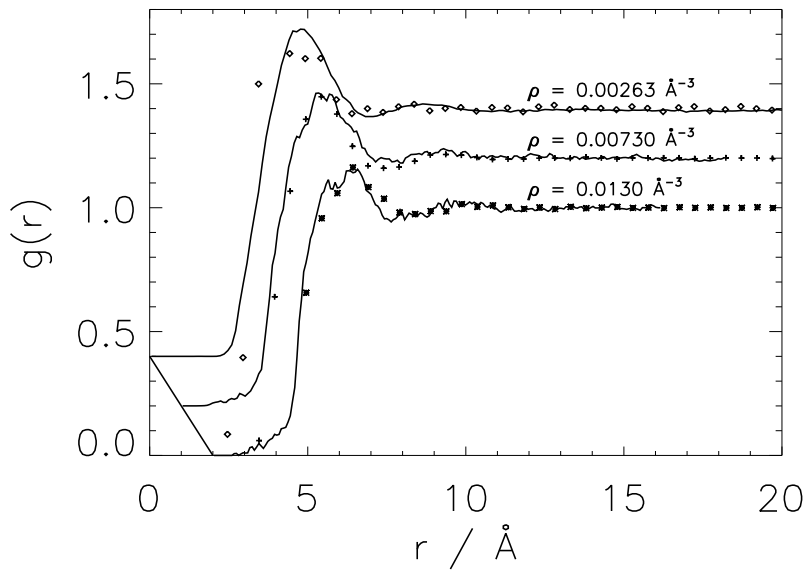}

\rule{0pt}{0pt}
\end{center}
\caption{\label{abb-h3cf-c-simexp-gr}
Comparison of $g^{(x)}(r)$ for H$_3$CF from the experiment with the result 
from a Monte-Carlo simulation using the site-site potentials derived for 
HCF$_3$. \\The symbols correspond to the experimental $g^{(x)}(r)$, the 
separation of the symbols in x-direction corresponds to the 
experimental resolution, the solid line to the simulation. The 
$g^{(x)}(r)$ correspond to different 
densities shown in the figure (state point B, D and E in 
Fig.~\ref{abb-h3cf-qiq}).}
\end{figure}

The HCF$_3$ molecular potential has been determined as a site-site potential,
whereas for H$_3$CF only the X-ray weighted pair distribution 
function $g(r)^{(x)}$ was determined, and thus an independent determination
of H$_3$CF site-site potentials was not possible. The question arises as
to whether the HCF$_3$ site-site potentials can also be used to describe the 
H$_3$CF structure, i.e. if these potentials have a general applicability 
to all fluorocarbons. Figure~\ref{abb-h3cf-c-simexp-gr} indicates that the 
site-site potentials are indeed transferable to a different molecular 
species.

\begin{figure}[tbhp]
\begin{center}
\includegraphics[scale=.6]{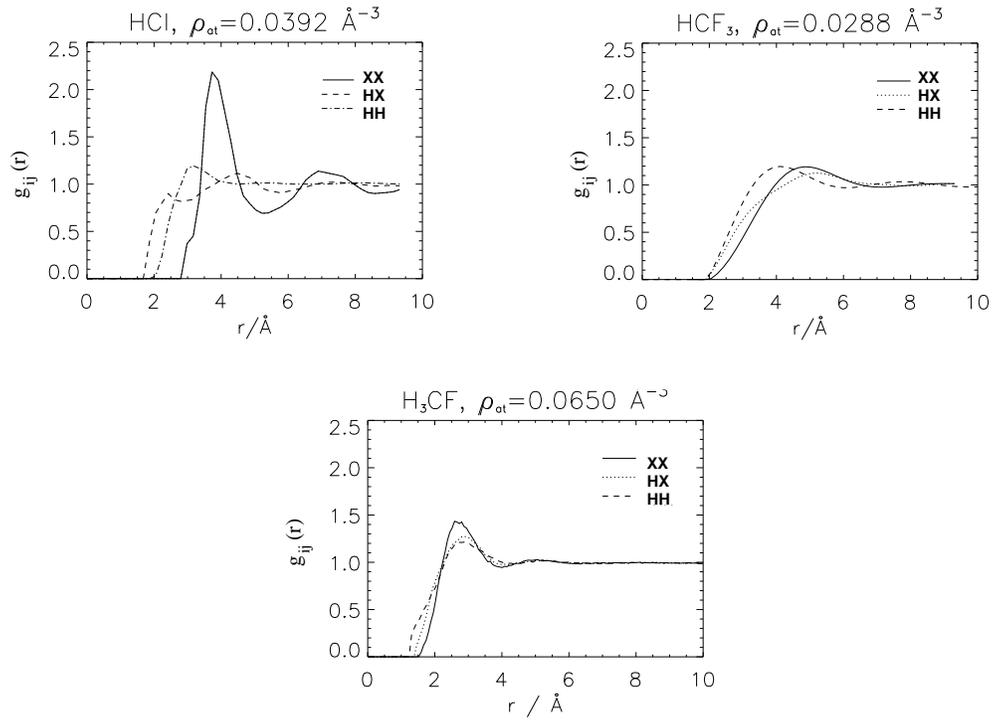}

\rule{0pt}{0pt}
\end{center}
\caption{Comparison of the HH, HX and XX CPPDF of HCF$_3$ and H$_3$CF 
with the PPDF of HCl\cite{And97}\label{abb-vgl-soper}.\\
The CPPDF for H$_3$CF are simulation results}
\end{figure}

The aim of the present work was to determine the influence of the molecular 
dipole on the orientational correlations between the molecules. 
Fig.~\ref{abb-vgl-soper} compares the HH, HX and XX pair distribution 
functions of fluid HCl \cite{And97}, HCF$_3$ and H$_3$CF.
 The three PPDF of HCl are quite structured and dissimilar 
while the corresponding CPPDF of both HCF$_3$ and H$_3$CF are much 
less structured and very similar. This behavior is an indication that 
there will be no strong 
preference for particular orientations in the fluorocarbons.
The most remarkable difference can be seen in the XX-(C)PPDF: $g_{ClCl}$ 
in HCl
has the highest maximum.

\begin{figure}[htbp]

\begin{minipage}{14cm}
a:\includegraphics[height=4.5cm,width=6.7cm]
{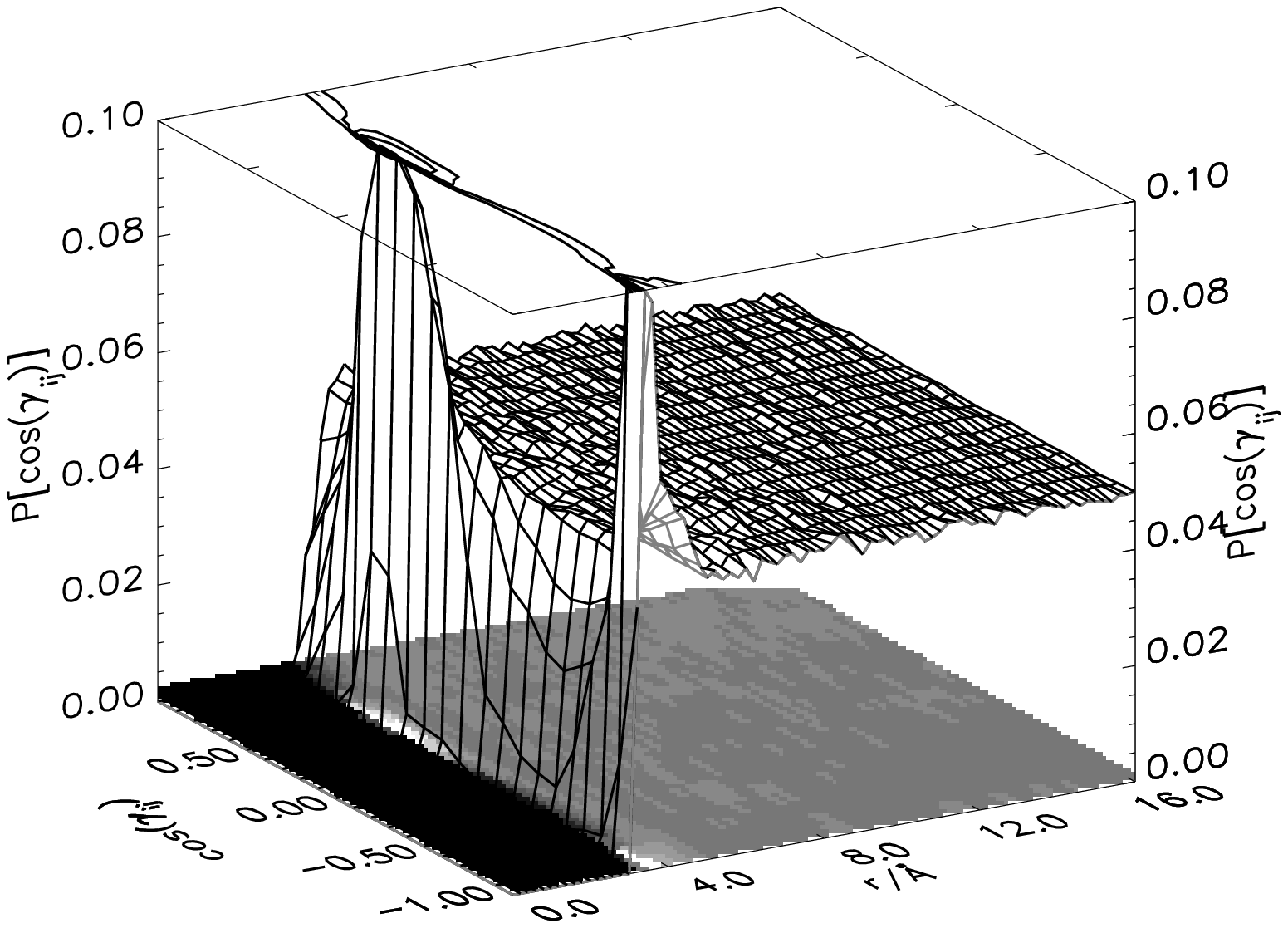}
b:
\includegraphics[height=4.5cm,width=6.7cm]
{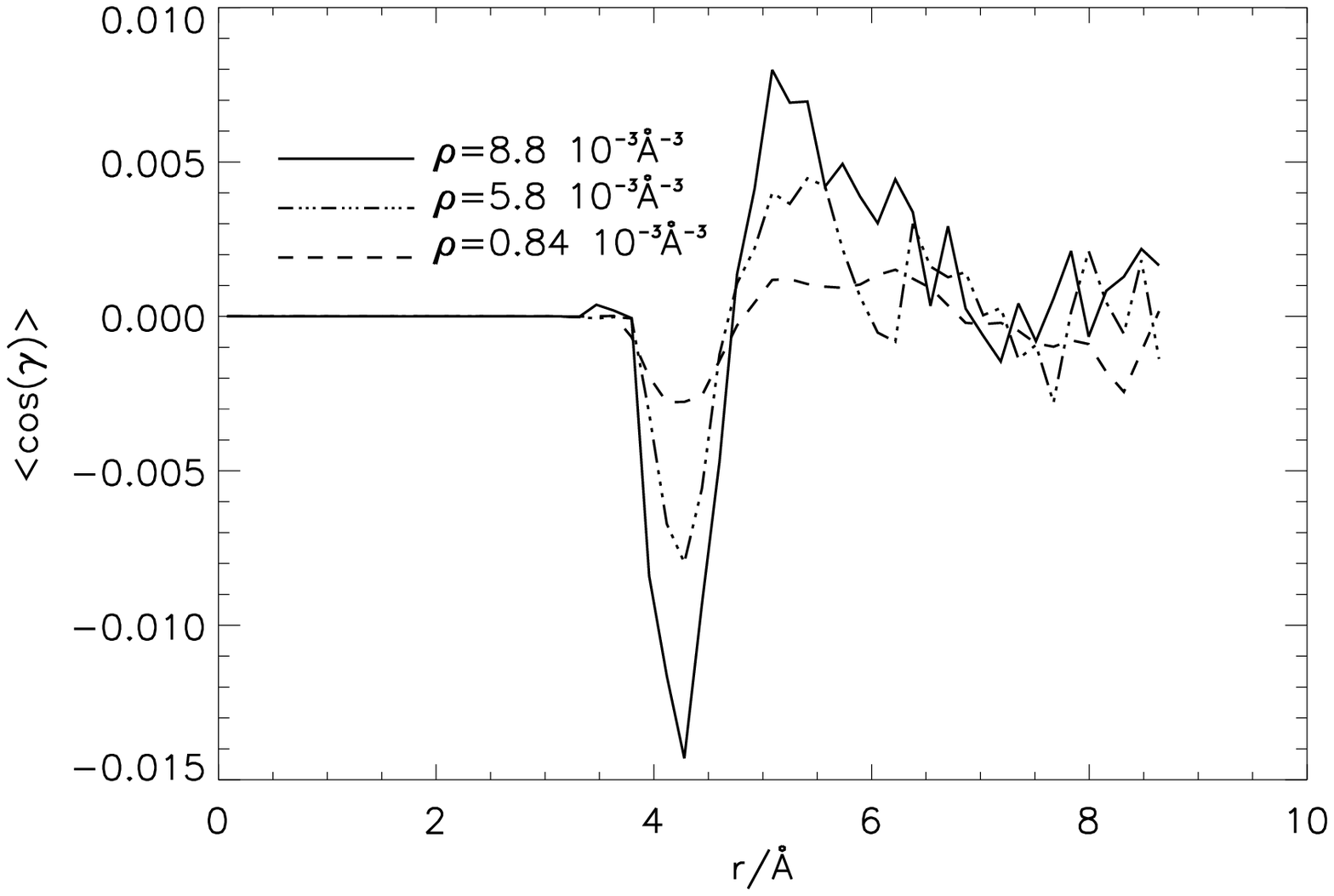}
\end{minipage}
\caption{a: Cosine distribution of the angle $\theta$ between 
the molecular dipoles of HCF$_3$ at
$\rho$=8.8 10${-3}$\AA$^{-3}$, \\b: 
Mean cosine of the angle $\theta$ {\it versus} the center of mass distance
at three different densities. \label{abb-cos-HCF3}}
\end{figure}

Fig.~\ref{abb-cos-HCF3} quantifies this qualitative statement and shows
$P(r_{COM},\cos(\theta))$, the relative probability of finding a second 
molecule at the center of mass distance
$r_{COM}$ in an orientation $\cos(\theta)$, where $\theta$ is the 
angle between the molecular dipoles. This figure is to be compared to 
figures 5 and 7 of Ref.~\cite{And97} and to figure 9 of Ref.~\cite{Wal}. 
While the first work determines the orientational correlation in fluid HCl 
via the EPMC formalism (truncated version of Eq.~\ref{eq-iter}) and finds
pronounced orientational correlations, the second is a Reverse Monte-Carlo 
study using a total neutron structure factor measurement of fluid CF$_4$ 
as input and  finds a  $P(r_{COM},\theta)$ very similar to 
fig.~\ref{abb-cos-HCF3}, structured only at very short distances.

Averaging the mean $<\cos(\theta)>_r$ via
\begin{equation}
g_{K} = 1+ \int_0^\infty N(r)<\cos(\theta)>_r dr 
\end{equation}
\begin{figure}[htbp]
a)\includegraphics[scale=.3]{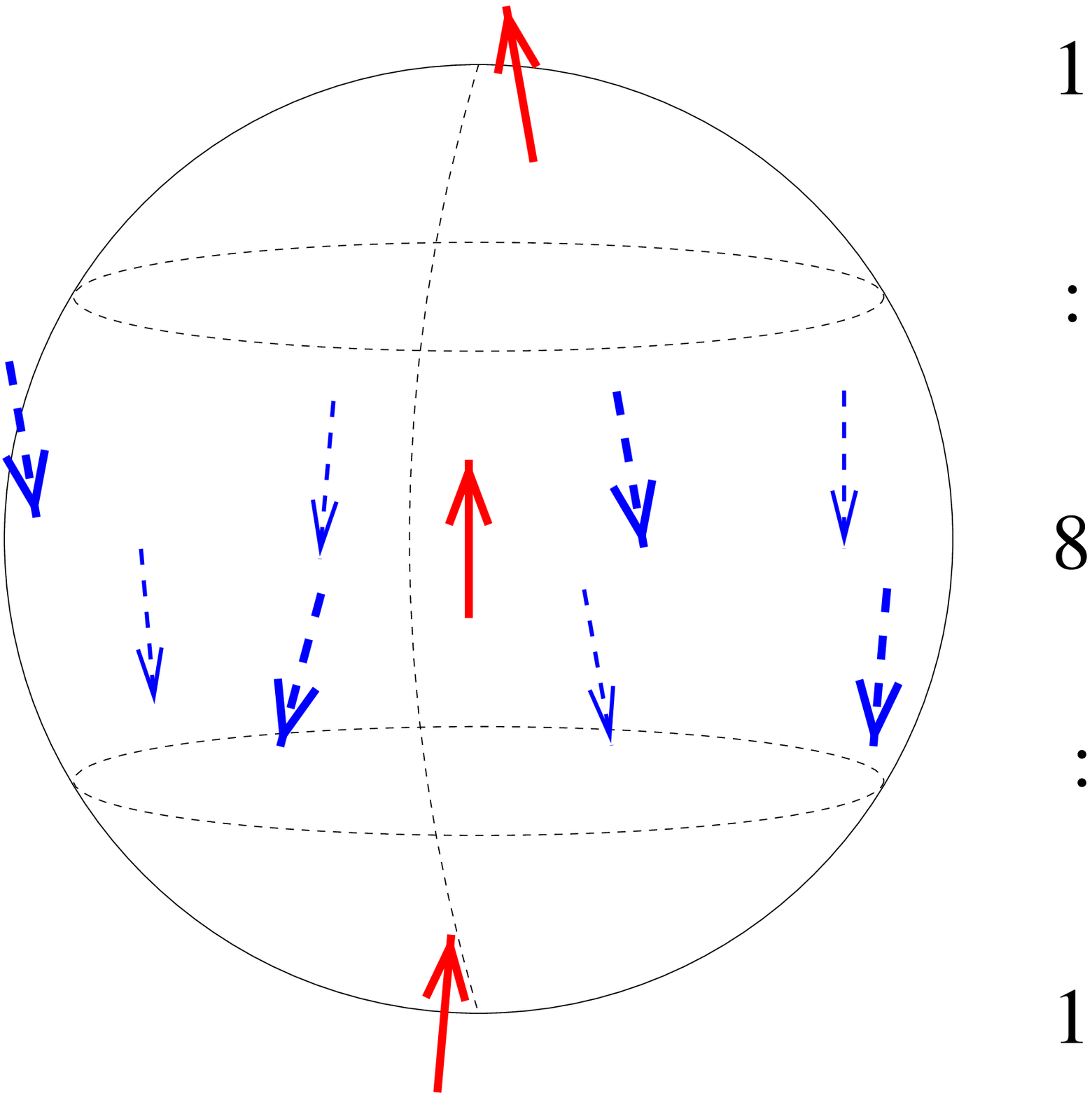}
b)\raisebox{-1.5cm}
{\includegraphics[scale=.6]{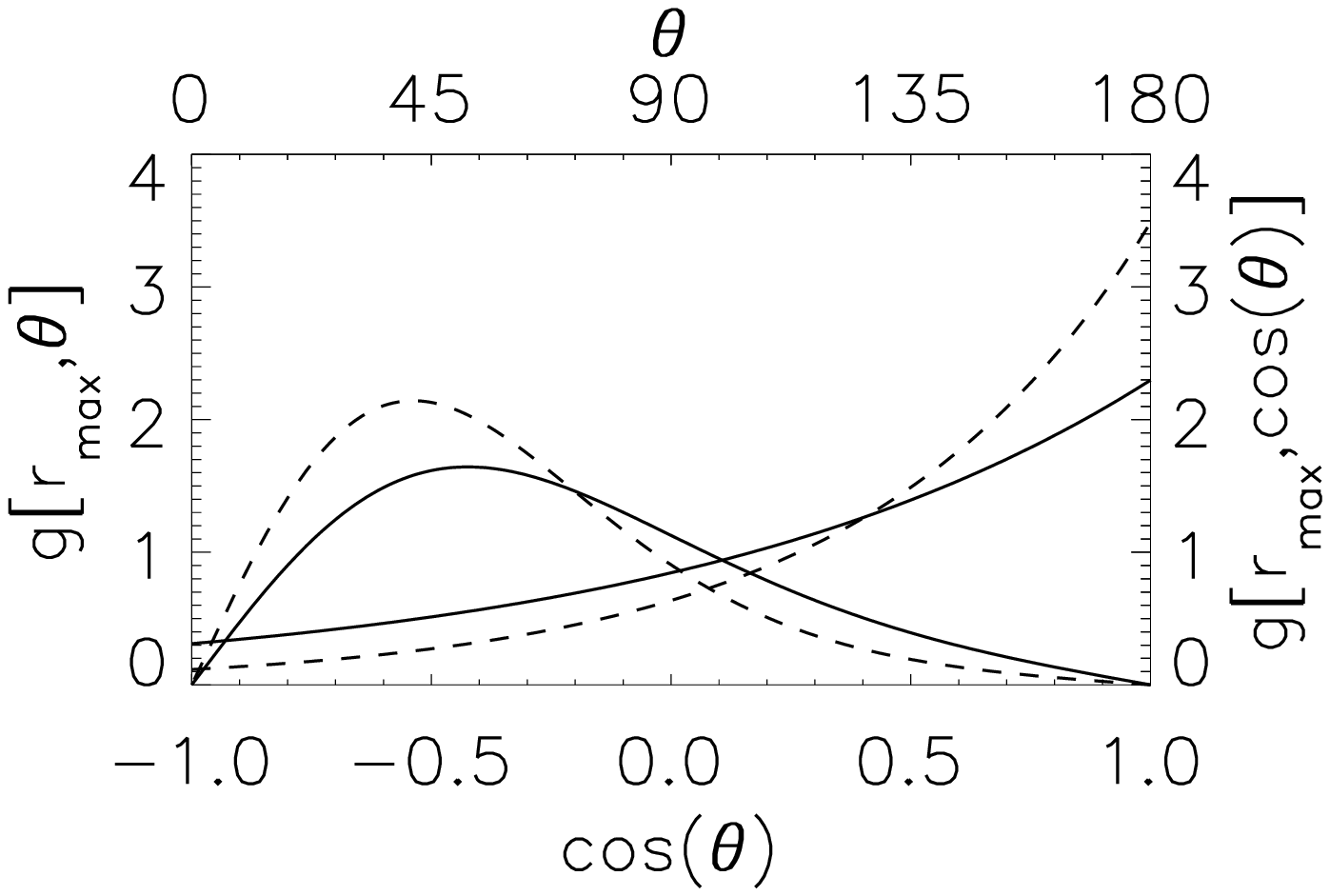}}
\caption {a) Simplified model of the relative orientation of dipolar 
molecules.
Explanation in the text. b) Cosine distribution and theta distribution
to be expected from two point dipoles in polar arrangement for
$p=1.65\cdot 3.336 \cdot 10^{-30}$\, Cm, $r_{max}$=5\AA\ (HCF$_3$, solid line)
and 
$p= 1.07\cdot 3.336\cdot10^{-30}$\, Cm, $r_{max}$=
3.6 \AA\
(HCl, broken line)}\label{over}
\end{figure}
with $N(r)= 4\pi \rho_{uc} r^2 g_{\rm COM}$,
to yield a Kirkwood g-factor leads to values very close to one 
(0.995 on average) in the entire density range investigated for both
HCF$_3$ and H$_3$CF. This is 
in agreement with advanced theories of the dielectric properties of 
these materials \cite{Sch89}.
 
The largely simplified model illustrated in Fig.~\ref{over}a can help to 
understand this behavior. At a distance of 5\AA, the maximum 
of the pair distribution
functions of HCF$_3$, 
the energy difference between parallel and antiparallel alignment 
of point dipoles of $1.65\cdot 3.336\cdot10^{-30}$\, Cm located at the 
center of mass 
is about 2kT for the polar positions and about
1kT for the equatorial positions. In a two level system  this would lead to
seven times more parallel than antiparallel orientation in polar positions 
and three times more antiparallel orientation in equatorial positions.
But the equatorial region is four times larger, leading to an almost 
complete cancelation of parallel and antiparallel orientations. 

For HCl a preference for a polar arrangement of molecules has been found
\cite{And97} (molecules directly 'below' or 'on top' of each other referred 
to the direction of the dipole). 
For these molecules in polar arrangement a strong 
preference for parallel orientations has been found - up to
17 times more parallel orientated dipoles than expected from a 
random distribution.  
The cosine distribution to be expected for point dipoles in a polar 
arrangement is:
\begin{equation}
P[r,\cos(\theta)]=\frac {\exp[k(r) \cos(\theta)]}{\int_{-1}^{1}\exp[k(r) 
\cos(\theta)] d\cos{\theta}}
\end{equation}
with $k(r)=2\cdot p^2/4\pi\epsilon r^3 $. 
This function is shown for $r=r_{max}$, with $r_{max}$ the first maximum 
of $g(r_{COM})$ in Fig.~\ref{over}b.
With $p=1.07\cdot 3.336 \cdot10^{-30}$\,Cm \cite{Gre} and $r_{max}=3.6$\,\AA\ 
thus only
3.5 time more parallel oriented dipoles as in a random distribution 
should be found. From a simple point dipole model a less pronounced
preference of parallel orientation as found by \cite{And97} is predicted 
for HCl, 
while on the other hand from this model a stronger
preference than actually found would be expected for both,
HCF$_3$ and H$_3$CF, probably due to the detailed molecular geometry
and specific site-site interactions.
These two effects, the cancelation of parallel and anti-parallel 
orientations and specific site-site interactions lead 
finally to the average behavior shown in Fig.~\ref{abb-cos-HCF3} very 
similar to fluid CF$_4$.

Hloucha {\it et al.} \cite{Dei} recently published a constant NPT Monte-Carlo 
simulation of liquid HCF$_3$  at subcritical temperatures. Their model 
is a rigid five site model, with  a Lennard-Jones contribution, 
partial charges at the sites and point dipoles at the center of mass, with
a constant and an induced contribution. With this model
and in the dense liquid more pronounced orientational correlations between 
neighboring molecules as in this work are found. The positive peak in 
$<\cos(\theta)>_r$ at $r \sim 4.5$\,\AA\  in Fig.~\ref{abb-cos-HCF3} 
is higher by a factor of two and the negative peak at close contact is 
missing. Hloucha {\it et al.} observed a decreasing trend in the 
orientational order with decreasing density which will tend to level out
the differences in orientational order in the range of densities investigated 
here. Although the detailed comparison is complicated by the difference in the
 range
of thermodynamic parameters investigated, this is supporting the point of 
view that the orientational ordering in HCF$_3$ is even less pronounced 
than predicted by a dipolar picture.

The spatial arrangement of neighboring molecules is illustrated in 
Fig.~\ref{HCF3-nice} which compares the crystal (a) --
 the positional parameters are 
taken from \cite{Tor96} -- and the fluid at liquid like densities(b). In the 
crystal each trifluoromethane molecule is surrounded by twelve neighboring 
molecules at nearly the same distance. Among these, two are oriented parallel
and two antiparallel, the remaining two times four molecules
in two different T-orientations,
that are orientations where the dipole moments are perpendicular to each other.
Fig.~\ref{HCF3-nice}b is a snap-shot of a simulation at 
$\rho_{uc}=0.0088$\AA$^{-3}$. 
Again, the twelve next-neighbors are shown. The orientation of these
molecules has been grouped into four classes having a $\cos(\theta)$ between
-1.0 and -0.5, -0.5 and 0.0, 0.0 and 0.5 and 0.5 and 1.0, respectively, 
with $\theta$ the
angle between the molecular dipoles. Again, quasi T-orientations
occur more often. In the fluid this snap-shot is only representative 
of course, the ensemble of structures leads to Fig.~\ref{abb-cos-HCF3}.
\begin{figure}[htb]
\caption{Relative orientation of next-neighbor molecules in 
crystalline HCF$_3$ (a) and dense fluid HCF$_3$ (b).\label{HCF3-nice}\\
The relative orientation to the central molecule is indicated with a 
color code.
In the crystal: Red (R): parallel, green (G) antiparallel, blue (B) and black
(b): different 
T-configurations. In the fluid: red $1.0>\cos(\theta)>0.5$, 
blue 0$.5>\cos(\theta)>0$, 
green $0>\cos(\theta)>-0.5$, black $-0.5>\cos(\theta)>-1.0$.}
\end{figure}
\section{Conclusion}

The density dependence of orientational correlations in fluid trifluoromethane
and methylfluoride has been studied by NVT-Monte-Carlo simulations using
effective site-site two body potentials derived via the Levesque-Weis-Reatto
inversion scheme from NDIS and hard X-ray diffraction data of fluid 
trifluoromethane. Advanced theories of the dielectric properties of these 
materials predict a Kirkwood-g-factor close to one in the entire density
range investigated here. The orientational correlations found in the 
simulation are in full agreement with this prediction. Detailed comparison 
with the orientational correlations found in fluid hydrogenchloride and 
tetrafluoromethane shows that the orientational correlation in HCF$_3$ and 
H$_3$CF are closer to tetrafluoromethane than to hydrogenchloride, although 
the interaction energy of the molecular dipoles is comparable to HCl in these
systems. This and the comparison with the simulation results for a 
dipolar model of HCF$_3$ suggests that in the fluorocarbons site specific
interaction results in weaker orientational correlations 
than predicted by a dipolar model while in fluid HCl the orientational
ordering is enhanced.


\section{Acknowledgment}
The assistance of P. Palleau and O. Koch   
during the neutron experiment and the help of A. Swiderski in constructing
the pressure cell is gratefully acknowledged.

\rule{0pt}{0pt}

\end{document}